\documentclass[conference,letterpaper]{IEEEtran}

\usepackage[utf8]{inputenc} 
\usepackage[T1]{fontenc}
\usepackage[cmex10]{amsmath}
\usepackage{amsfonts}
\usepackage{bm}
\usepackage{url}
\usepackage{ifthen}
\usepackage{cite}
\usepackage{cleveref}
\usepackage{xcolor}
\usepackage{tikz}
\usepackage{subfig}
\usetikzlibrary{shapes,arrows,positioning,fit,backgrounds,calc}

\DeclareMathOperator{\arctanh}{arctanh}
\DeclareMathOperator{\erf}{erf}
\DeclareMathOperator{\sgn}{sgn}
\DeclareMathOperator{\DiffAttention}{DiffAttention}
\DeclareMathOperator{\softmax}{softmax}

\begin{document}
\title{Interplay Between Belief Propagation and Transformer: Differential-Attention Message Passing Transformer} 


\author{%
  \IEEEauthorblockN{Chin Wa (Ken) Lau*}
  \IEEEauthorblockA{
  The Chinese University of Hong Kong \\
  kenlau@ie.cuhk.edu.hk}
  
  \and
  \IEEEauthorblockN{Xiang Shi*, Ziyan Zheng}
  \IEEEauthorblockA{Tsinghua University\\
  \{shix22,zhengzy19\}@mails.tsinghua.edu.cn \\
  *These authors contributed equally.}
  \and
  \IEEEauthorblockN{Haiwen Cao, Nian Guo}
  \IEEEauthorblockA{Huawei Technologies Co., Ltd. \\
  \{cao.haiwen,guonian4\}@huawei.com}
}

\maketitle


\begin{abstract}
    Transformer-based neural decoders have emerged as a promising approach to error correction coding, combining data-driven adaptability with efficient modeling of long-range dependencies. This paper presents a novel decoder architecture that integrates classical belief propagation principles with transformer designs. We introduce a differentiable syndrome loss function leveraging global codebook structure and a differential-attention mechanism optimizing bit and syndrome embedding interactions. Experimental results demonstrate consistent performance improvements over existing transformer-based decoders, with our approach surpassing traditional belief propagation decoders for short-to-medium length LDPC codes.
 \end{abstract}

\section{Introduction}
Classical error-correcting codes (ECC) have long been the cornerstone of reliable digital communications. While traditional decoders like belief propagation (BP) and successive cancellation list (SCL) have served well, neural decoders show promising potential to learn and adapt to channel characteristics. This data-driven approach has led to extensive exploration of various architectures \cite{mao24}, from feedforward neural networks (FFNNs) \cite{gch17, slk18} and convolutional neural networks (CNNs) \cite{lzj18, zcz20} to recurrent neural networks (RNNs) \cite{kjr18, bck18}, aiming to achieve superior performance compared to conventional methods.

A fundamental challenge in designing neural decoders lies in effectively capturing long-range dependencies among code-bits while managing reasonable training complexity through efficient utilization of code structure. Traditional approaches using fully connected architectures encounter significant scalability limitations \cite{gch17}, particularly as code lengths increase. This dual challenge of managing computational complexity and effectively incorporating code structure has been a critical bottleneck in developing practical neural decoders.

The emergence of transformer architectures presents a promising solution to these challenges. Their remarkable success in language models stems from an inherent ability to model complex, long-range correlations through attention mechanisms. This capability, when combined with domain-specific knowledge of codebook structure, has enabled transformer-based decoders to achieve performance comparable to classical BP decoders \cite{chw22, pkk24}.

The synergy between classical coding theory and transformer architectures offers a powerful framework for decoder design. Recent works have demonstrated this potential, with \cite{chw22} incorporating a parity check matrix to guide masked self-attention for reducing training complexity, and \cite{pkk24} achieving breakthrough performance through iterative updates of signal magnitudes and hard syndromes. This integration of domain knowledge with modern neural architectures represents a promising direction for advancing decoder performance.

Building on these foundations, we introduce several key innovations to enhance transformer-based decoders. Our contributions include a novel syndrome loss function that leverages global codebook structure (\Cref{sec:syn-loss}), an improved architecture that incorporates message-passing principles (\Cref{sec:input-output}), and a differential attention mechanism (\Cref{sec:diff-attn}) that refines attention patterns. Comprehensive experimental results in \Cref{sec:result} demonstrate the effectiveness of these enhancements compared to existing approaches.

\section{Background}
\subsection{Notations and Channel Model}
In this work, we focus on forward-error correction codes for a binary codebook $\mathcal{C}$ transmitted over an additive white Gaussian noise (AWGN) channel using binary phase-shift keying (BPSK) modulation.

The encoder maps a $k$-bit message $\mathbf{u} \in \text{GF}(2)^k$ to an $n$-bit codeword $\mathbf{c} = \mathbf{u} \mathbf{G} \in \text{GF}(2)^n$ using a generator matrix $\mathbf{G} \in \text{GF}(2)^{k \times n}$. The encoder then modulates the codeword into a bipolar signal $\mathbf{x} = 1 - 2\mathbf{c}$.

The decoder receives a corrupted signal $\mathbf{y} = \mathbf{x} + \mathbf{z}$, where $\mathbf{z}$ is a noise vector generated independently from a normal distribution $\mathcal{N}(0, \sigma^2)$, and $\sigma^2$ is the known channel noise variance. We adopt the following relationship between the channel noise variance and the normalized signal-to-noise ratio (SNR) $E_b/N_0$:
\begin{align}
\frac{E_b}{N_0} = 10 \log_{10} \frac{1}{2\sigma^2 R},
\end{align}
where $R = k/n$ denotes the code rate.

For BPSK modulation over an AWGN channel, which forms a binary-input symmetric-output (BISO) channel, we apply an alternative formulation from \cite[Lemma 1]{riu01} to model $\mathbf{y} = \mathbf{x} \cdot \tilde{\mathbf{z}}$, where $\tilde{\mathbf{z}}$ represents independently generated multiplicative noise.

The decoder utilizes the log-likelihood ratio (LLR) $\bm{y}^\dagger = 2\mathbf{y}/\sigma^2$ to generate a soft decision $\mathbf{c}^\dagger \in \mathbb{R}^n$ on the codewords, and the hard decision is given by $\hat{\mathbf{c}} = 0.5(1 - \sgn(\mathbf{c}^\dagger))$.

To introduce a priori knowledge into the neural decoder architecture and its loss function, we utilize the parity-check matrix $\mathbf{H} \in \text{GF}(2)^{(n-k) \times n}$ of the code $C$ and its corresponding Tanner graph $\mathcal{G}$. We define $\{v_i\}_{i=1}^n$ and $\{c_j\}_{j=1}^{n-k}$ as the sets of variable nodes and check nodes, respectively, and we use the boundary symbol $\partial$ to denote the neighborhood of a given vertex.

\subsection{Soft Syndrome}
While no standardized definition of soft syndrome exists in coding theory, this concept has emerged as a valuable tool in the design of neural decoders. Soft syndromes provide essential side information that enables neural decoders to effectively utilize prior knowledge from syndrome decoding. Furthermore, the continuous nature of soft syndromes, as opposed to the discrete values of conventional hard syndromes, facilitates more efficient training of neural networks through gradient-based optimization.

The formulation of soft syndromes is predominantly inspired by message-passing techniques from iterative decoding algorithms \cite{lug18, tec20}, particularly the sum-product algorithm (SPA) and min-sum algorithm (MSA). Taking inspiration from the SPA's check node processing, we define the soft syndrome $\mathbf{s}^\dagger$ as a function of the log-likelihood ratio (LLR) of the received signal $\mathbf{y}^\dagger$:
\begin{align}
    s_j^\dagger = 2 \arctanh\left( \prod_{i : v_i\in \partial c_j} \tanh\left(\frac{y_i^\dagger}{2}\right)\right).
\end{align}

The conventional hard syndrome $\mathbf{s}$ can be recovered from the soft syndrome through the following transformation: $\mathbf{s} = 0.5 (1 - \sgn(\mathbf{s}^\dagger))$.

\subsection{Transformer-based Neural Decoders}
\begin{figure}[htbp]
    \centering
    \begin{center}
\begin{tikzpicture}[
    node distance=2cm,
    box/.style={draw, rounded corners=2pt, minimum width=1.5cm},
    block/.style={draw, rounded corners=2pt, minimum width=2cm},
    matrixblock/.style={draw, rounded corners=2pt, minimum width=0.5cm, minimum height=0.5cm},
    bigbox/.style={draw=blue!20, dashed, very thick, rounded corners, inner sep=0cm},
    arr/.style={->, thick, >=latex},
    xscale=0.8, yscale=0.7
]

\node[bigbox, minimum width=6cm, minimum height=3cm] (postproc) at (0,7.5) {};

  \node at (0,10) {\textbf{Post-processing}};
\node[box] (z) at (0,9) {$\mathbf{\bar{z}}$};
\node[block, draw=red!40, fill=red!10] (fc) at (0,7.75) {Fully Connected};
\node[matrixblock, draw=orange!40, fill=orange!10] (H) at (0,6.75) {\textbf{H}};
\node[block, draw=red!40, fill=red!10] (proj1) at (-2,6) {Projection};
\node[block, draw=red!40, fill=red!10] (proj2) at (2,6) {Projection};
\node at (-3.25,7.25) {$\{\varphi_i \in \mathbb{R}\}_{i = 1}^n$};
\node at (3.3,7.25) {$\{\psi_j \in \mathbb{R}\}_{j = 1}^{n - k}$};

\node[draw=blue!40, fill=blue!10, rounded corners, minimum width=5cm, minimum height=0.5cm] (decoder) at (0,4) {};
\node at (0,4) {Decoder Layer $\times N$};
\node[bigbox, minimum width=6cm, minimum height=2.2cm] (preproc) at (0,0.5) {};
\node at (0,2.5) {\textbf{Pre-processing}};

\node[box, draw=green!40, fill=green!10] (llr) at (-2,0) {$|\mathbf{y}^\dagger|$};
\node[box, draw=green!40, fill=green!10] (syn) at (2,0) {$\mathbf{s}^\dagger$};
\node[box, draw=green!40, fill=green!10] (embed) at (0,1.5) {Embedding};  
\node at (-2,-0.7) {Absolute LLR};
\node at (2,-0.7) {Soft Syndrome};
\node at (-3.5,2.75) {$\{\bm{\phi}_i \in \mathbb{R}^d\}_{i = 1}^n$};
\node at (3.5,2.75) {$\{\bm{\psi}_j \in \mathbb{R}^d\}_{j = 1}^{n - k}$};

\node[circle,draw,inner sep=1pt] (mul1) at (-2,1.5) {$\times$};
\node[circle,draw,inner sep=1pt] (mul2) at (2,1.5) {$\times$};

\draw[arr] (fc) -- (z);
\draw[arr] (H) -- (fc);
\draw[arr] (proj1) |- (fc.west);
\draw[arr] (proj2) |- (fc.east);
\draw[arr] let \p1 = (decoder.north) in (-2,\y1) -- (proj1);
\draw[arr] let \p1 = (decoder.north) in (2,\y1) -- (proj2);
\draw[arr] (llr) -- (mul1);
\draw[arr] (syn) -- (mul2);
\draw[arr] let \p1 = (decoder.south) in (mul1) -- (-2,\y1);
\draw[arr] let \p1 = (decoder.south) in (mul2) -- (2,\y1);
\draw[arr] (embed) -- (mul1);
\draw[arr] (embed) -- (mul2);
\end{tikzpicture}
\end{center}
    \caption{Transformer-based Neural Decoder Architecture}
    \label{fig:trans-arch}
\end{figure}
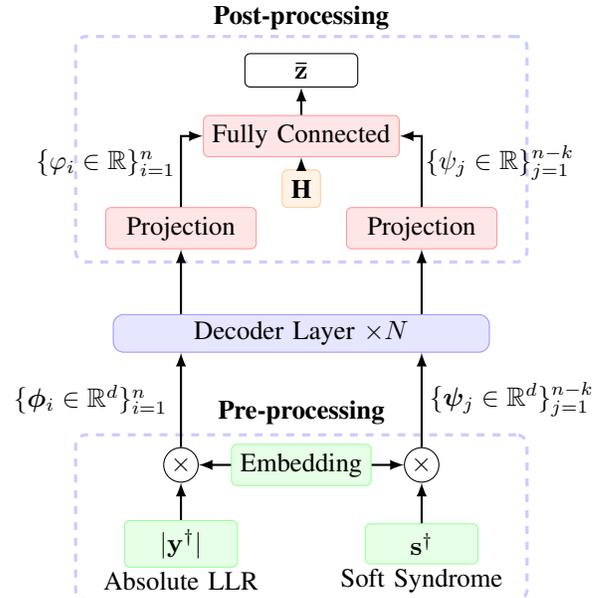

We provide a high-level overview of our transformer-based neural decoder architecture, as illustrated in \Cref{fig:trans-arch}, to contextualize our contributions.

The architecture consists of three primary components: pre-processing, transformer decoder, and post-processing modules. The pre-processing module embeds a sequence of input scalars $\{t_i\}$ into $d$-dimensional vectors $\{\bm{\phi}_i\}$, commonly referred to as embeddings. The basic embedding operation is given by $\bm{\phi}_i = t_i\mathbf{w}_i$, where $\{\mathbf{w}_i\}$ are learnable parameter vectors.

The post-processing module performs a complementary operation, transforming the sequence of $d$-dimensional embedding vectors $\{\bm{\phi}_i\}$ into scalar values $\{\varphi_i\}$, which are then used to compute the soft decision $\mathbf{c}^\dagger$ on the codewords. In \Cref{sec:input-output}, we present enhancements to both pre-processing and post-processing modules inspired by message passing algorithms.

The transformer decoder module consists of multiple decoding blocks, which can be viewed as an unrolled version of an iterative algorithm. Each decoding block comprises an attention module and a feed-forward module. The attention module learns the correlations between input embeddings by computing weighted interactions among them, while the feed-forward neural network (FFNN) applies nonlinear transformations to the embedding vectors to refine their representation in the latent space. In \Cref{sec:diff-attn}, we adopt a novel attention mechanism from \cite{ydx24} that effectively captures the relationship between received signals and soft syndrome by leveraging the structure induced by the parity-check matrix.

\section{Integrating Belief Propagation in Attention}
\subsection{Syndrome-based Loss}
\label{sec:syn-loss}
Most neural channel decoders employ binary cross entropy $\mathcal{L}_\text{transport}$ as their loss function during supervised learning. This approach implicitly treats forward error correction as a series of binary classification problems, focusing on optimizing the bit error rate. To enhance the block error rate performance, we introduce a complementary loss term $\mathcal{L}_\text{validation}$ that leverages the Tanner graph structure. This validation loss serves as a regularization term by measuring the syndrome validity of estimated codewords.

The total loss function combines the traditional transport loss with our validation loss:
\begin{align}
    \mathcal{L} &= \mathcal{L}_\text{transport} + \lambda_\text{multi-loss} \mathcal{L}_\text{validation},
\end{align}
\begin{align}
    \mathcal{L}_\text{transport} &= \frac{1}{n} \sum_i -\Pr(C_i = 0) \log \Pr(\hat{C}_i = 0) \\
    &\quad - \Pr(C_i = 1) \log \Pr(\hat{C}_i = 1), \nonumber \\
    \mathcal{L}_\text{validation} &= \frac{1}{n - k} \sum_j -\log \Pr(S_j = 0).
\end{align}
The validation loss is computed as the binary cross entropy between the estimated syndrome and the all-zero syndrome, since valid codewords must satisfy all parity-check equations.
Previous works \cite{lug18, tec20} proposed syndrome-based losses inspired by the min-sum algorithm. However, their soft syndrome calculations are non-differentiable, which limits the effectiveness of gradient-based training. We address this limitation by introducing a novel syndrome loss based on mean-field approximation, which provides both theoretical foundations and differentiability.

To compute $\Pr(S_j = 0)$, we apply the mean-field approximation \cite{yfw05}, which assumes independence:
\begin{align}
\Pr(\hat{C}_1 = \hat{c}_1, \dots, \hat{C}_n = \hat{c}_n) = \prod_{i = 1}^n \Pr(\hat{C}_i = \hat{c}_i).
\end{align}

For a full-rank parity-check matrix, this leads to independent syndrome components:
\begin{align}
    \Pr(S_1 = s_1, \dots, S_{n - k} = s_{n - k}) = \prod_{j = 1} \Pr(S_j = s_j).
\end{align}

Leveraging the binary nature of codewords and the mean-field approximation, the probability of satisfying check node $c_j$ can be computed using \cite[Lemma 1]{gal62}:
{\small
\begin{align}
    \Pr(S_j = 0) = \frac{1}{2} \left(1 + \prod_{i: v_i \in \partial c_j} (\Pr(\hat{C}_i = 0) - \Pr(\hat{C}_i = 1)) \right). 
\end{align}}

This formulation provides an efficient and differentiable approximation of the syndrome loss.

\subsection{Enhanced Input Representation and Embedding Aggregation}
\label{sec:input-output}

Theoretically, if we could train a perfect neural decoder, any sufficient statistics would be valid for the input and output representations in decoding. However, due to practical constraints such as memory and computational limitations that restrict training to only a fraction of possible codewords, and the inherent limitations of model architectures, the choice of input and output representations significantly impacts the decoding performance of neural decoders.

Our experiments demonstrate improved performance using absolute LLR values $|\bm{y}^\dagger|$ and soft syndrome $\mathbf{s}^\dagger$ as inputs to the pre-processing module. According to \cite[Theorem 1]{bck18}, this choice preserves optimal performance. The model predicts the LLR of multiplicative noise $\tilde{\mathbf{z}}^\dagger = f_\theta(|\bm{y}^\dagger|, \mathbf{s}^\dagger)$ in the post-processing module, differing from prior approaches \cite{chw22, pkk24} that rely on absolute input signals $|\mathbf{y}|$ and hard-decision syndromes $\mathbf{s}$. The soft decision of codewords is computed as $\mathbf{c}^\dagger = \sgn(\mathbf{y}) \cdot \tilde{\mathbf{z}}^\dagger$. In our network architecture, we encode input features as bit embeddings $\bm{\phi}_i = |y_i^\dagger| \mathbf{w}_i$ and syndrome embeddings $\bm{\psi}_j = s_j^\dagger \tilde{\mathbf{w}}_j$, where $\mathbf{w}_i$ and $\tilde{\mathbf{w}}_j$ are learnable embedding vectors for the respective embeddings.

The advantage of this approach can be explained through an information-theoretic perspective when comparing the estimation of multiplicative noise versus direct codeword estimation. To enhance neural decoder performance, it is beneficial to minimize the entropy of the learnable components' output. The entropy of hard-decided codewords is $H(\hat{C}_1, \dots, \hat{C}_n) = \log_2 k$, while for hard-decision multiplicative noise it is $H(\tilde{Z}_1, \dots, \tilde{Z}_n) \approx n h_b(p)$, where $p = \frac{1}{2} - \frac{1}{2} \erf \left( \frac{1}{\sqrt{2} \sigma}\right)$ represents the bit-flipping probability in the binary input AWGN channel and $h_b$ is the discrete binary entropy. In practical SNR ranges, the multiplicative noise entropy is significantly lower than the codeword entropy, suggesting why estimating multiplicative noise leads to better decoding performance.

Furthermore, we enhance the final aggregation between bit embeddings and syndrome embeddings, as illustrated in \Cref{fig:trans-arch}. While previous transformer decoders \cite{chw22, pkk24} obtain the predicted multiplicative noise $\tilde{\mathbf{z}}$ through a simple fully-connected layer, our architecture leverages the structure of the code's Tanner graph. Specifically, we restrict the embedding aggregation patterns to follow the connectivity defined by the parity-check matrix's induced Tanner graph, leading to improved decoding performance.

\subsection{Tanner-graph Differential Attention}
\label{sec:diff-attn}
Recent work in neural decoding has shown that cross-attention between magnitude and syndrome embeddings outperforms full-attention mechanisms \cite{pkk24, chw22}. However, our analysis reveals a fundamental limitation: the background attention scores remain comparable to those between error-associated magnitude and syndrome embeddings. This issue stems from the softmax normalization constraint—even when a query embedding has minimal correlation with key embeddings, the neural decoder must distribute attention weights to sum to 1, potentially masking true error patterns in the received codewords.

To address this limitation, we introduce a differential cross-attention mechanism inspired by \cite{ydx24}. Our approach eliminates background attention noise while maintaining computational efficiency without additional parameters. The mechanism is defined as:
\begin{align}
    &\DiffAttention(Q, K, V; \mathbf{M}) := \\
    &\  \left(\softmax \left(\frac{Q K^T}{\sqrt{d}} + \psi(\mathbf{M})\right) - \lambda_\text{diff} \softmax \left(\frac{Q K^T}{\sqrt{d}}\right)\right)V, \nonumber
\end{align}
where $Q$, $K$, and $V$ are the query, key, and value matrices, $d$ is the input embedding dimension, and $\mathbf{M}$ is the mask matrix. The masking function $\psi$ is:
\begin{align}
    [\psi(\mathbf{M})]_{i, j} := \begin{cases}
        0 & \text{if $[\mathbf{M}]_{i, j} = 1$,}\\
        -\infty & \text{if $[\mathbf{M}]_{i, j} = 0$.}\\
    \end{cases}
\end{align}
This formulation captures desired attention scores through the first term while subtracting background noise via the second term. By reusing attention scores ($Q K^T$), we preserve computational efficiency while filtering spurious attention patterns.

\begin{figure}[htbp]
    \centering
    \begin{tikzpicture}[
    node distance=2cm,
    box/.style={draw, rounded corners=2pt, minimum width=2cm, minimum height=1pt},
    diffbox/.style={draw, fill=green!10, rounded corners=2pt, minimum width=1.5cm, minimum height=1pt},
    ffnnbox/.style={draw, fill=orange!10, rounded corners=2pt, minimum width=1.5cm, minimum height=1pt},
    normbox/.style={draw, rounded corners=2pt, minimum width=1.5cm, minimum height=1pt},
    wbox/.style={draw=orange!40, fill=orange!10, rounded corners=2pt},
    hbox/.style={draw, fill=yellow!10, rounded corners=2pt},
    bigbox/.style={draw=blue!20, fill=blue!5, rounded corners, inner sep=0.3cm},
    arr/.style={->, thick, >=latex},
    xscale=0.85, yscale=0.8
]

\node[bigbox, minimum width=8.5cm, minimum height=7cm] (decoderlayer) at (0,2.5) {};
\node[font=\bfseries] at (0,6.5) {Decoder Layer};

\node[normbox] (norm1) at (-2.75,0) {Norm};
\node[diffbox] (diff1) at (-2.75,2) {DiffAttn};
\node[ffnnbox] (ffnn1) at (-2.75,3.75) {FFNN};
\node[normbox] (norm2) at (-2.75,4.75) {Norm};
\node[hbox] (h1) at (-4.25,2) {\textbf{H}};

\node[normbox] (norm3) at (2.75,0) {Norm};
\node[diffbox] (diff2) at (2.75,2) {DiffAttn};
\node[ffnnbox] (ffnn2) at (2.75,3.75) {FFNN};
\node[normbox] (norm4) at (2.75,4.75) {Norm};
\node[hbox] (h2) at (4.25,2) {\textbf{H}};

\node[wbox] (wq1) at (-2.75,1) {$W_Q$};
\node[wbox] (wk1) at (-1,2.5) {$W_K$};
\node[wbox] (wv1) at (-1,1.5) {$W_V$};

\node[wbox] (wq2) at (2.75,1) {$W_Q$};
\node[wbox] (wk2) at (1,2.5) {$W_K$};
\node[wbox] (wv2) at (1,1.5) {$W_V$};

\node[circle,draw,inner sep=0.5pt] (add1) at (-2.75,2.75) {$+$};
\node[circle,draw,inner sep=0.5pt] (add2) at (-2.75,5.5) {$+$};
\node[circle,draw,inner sep=0.5pt] (add3) at (2.75,2.75) {$+$};
\node[circle,draw,inner sep=0.5pt] (add4) at (2.75,5.5) {$+$};

\node at (-2.75,-1.25) {Bit Embeddings};
\node at (2.75,-1.25) {Syndrome Embeddings};

\draw[arr] (norm1) -- (wq1);
\draw[arr] (wq1) -- (diff1);
\draw[arr] (diff1) -- (add1);
\draw[arr] (add1) -- (ffnn1);
\draw[arr] (ffnn1) -- (norm2);
\draw[arr] (norm2) -- (add2);
\draw[arr] (wq1) -- (diff1);
\draw[arr] (wk1) -- (diff1);
\draw[arr] (wv1) -- (diff1);
\draw[arr] (h1) -- (diff1);

\draw[arr] let \p1 = (h1.west) in (-2.75,-0.75) -- (\x1-5,-0.75) |- (add1);
\draw[arr] let \p1 = (h1.west) in (-2.75,3.1) -- (\x1-5,3.1) |- (add2);

\draw[arr] (-2.75, -1) -- (norm1);
\draw[arr] (add2) -- (-2.75, 6.25);
\draw[arr] (2.75,-0.75) -- (-0.1,-0.75) |- (wv1);
\draw[arr] (2.75,-0.75) -- (-0.1,-0.75) |- (wk1);

\draw[arr] (norm3) -- (wq2);
\draw[arr] (wq2) -- (diff2);
\draw[arr] (diff2) -- (add3);
\draw[arr] (add3) -- (ffnn2);
\draw[arr] (ffnn2) -- (norm4);
\draw[arr] (norm4) -- (add4);
\draw[arr] (wq2) -- (diff2);
\draw[arr] (wk2) -- (diff2);
\draw[arr] (wv2) -- (diff2);
\draw[arr] (h2) -- (diff2);

\draw[arr] let \p1 = (h2.east) in (2.75,-0.75) -- (\x1+5,-0.75) |- (add3);
\draw[arr] let \p1 = (h2.east) in (2.75,3.1) -- (\x1+5,3.1) |- (add4);

\draw[arr] (2.75, -1) -- (norm3);
\draw[arr] (add4) -- (2.75, 6.25);
\draw[arr] (-2.75,5.9) -- (0.1,5.9) |- (wv2);
\draw[arr] (-2.75,5.9) -- (0.1,5.9) |- (wk2);
\end{tikzpicture}
    \caption{Decoder Layer with Differential Cross-Attention}
    \label{fig:diff-attn}
\end{figure}
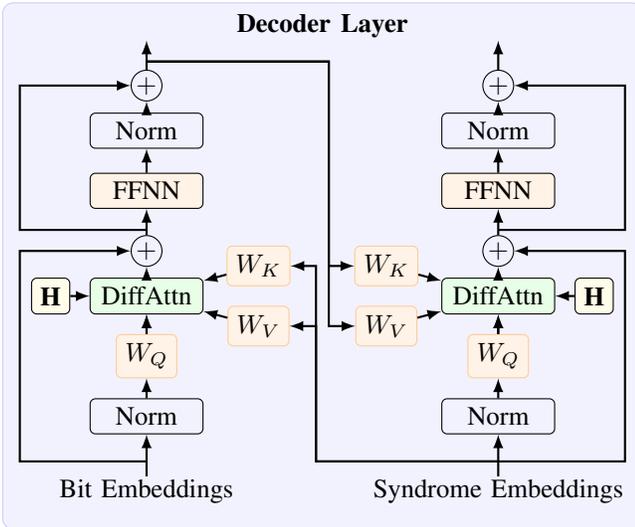

We integrate this differential attention into the iterative update framework of \cite{pkk24}, which treats bit embeddings $\{\bm{\phi}_i\}_{i = 1}^n$ and syndrome embeddings $\{\bm{\psi_j}\}_{j = 1}^{n - k}$ as multimodal data (\Cref{fig:diff-attn}). The update process occurs in two half-iterations:

The first iteration constructs attention matrices as follows:
\begin{itemize}
    \item Query matrices from bit embeddings: $Q = [\bm{\phi}_1; \dots; \bm{\phi}_n] W^Q$
    \item Key and value matrices from syndrome embeddings: 
        \begin{align}
            K &= [\bm{\psi}_1; \dots; \bm{\psi}_{n - k}] W^K \\
            V &= [\bm{\psi}_1; \dots; \bm{\psi}_{n - k}] W^V
        \end{align}
\end{itemize}
where $W^Q$, $W^K$, and $W^V$ are learnable projection matrices. The parity-check matrix $\mathbf{H}^T$ serves as the masking matrix. In the second half-iteration, we swap bit and syndrome embedding roles, using $\mathbf{H}$ as the mask. This structure implements message-passing while preserving the parity-check matrix's structural information.

\section{Numerical Results}
\label{sec:result}
\subsection{Weight Sharing and Training Strategy}
Building upon our differential attention mechanism, we implement strategic parameter sharing to enhance training efficiency while maintaining decoder performance. Our optimization strategy addresses two key aspects of the architecture: the differential cross-attention module and the bit-syndrome interactions.

In the differential cross-attention module, we implement weight sharing across projection matrices in different attention modules. This design is motivated by the fundamental requirement that attention scores before and after masking must be comparable within consistent latent spaces. By sharing weights between these components, we ensure that the background and masked attention patterns operate in the same dimensional space, enabling effective noise suppression as described in \Cref{sec:diff-attn}.

For bit-syndrome interactions, our weight sharing strategy extends to both attention modules and feedforward layers across the two half-iterations of the update process. This design is inspired by \cite{pkk24}, which suggests the potential for interaction between bit embeddings and syndrome embeddings in similar latent spaces. Consequently, sharing parameters between the first and second half-iterations maintains representational consistency while reducing the model's parameter count.

To enhance gradient flow and training dynamics, we adopt Gaussian Error Linear Units (GELU) \cite{heg16} in the feedforward neural networks, replacing traditional Rectified Linear Units (ReLU). This substitution aligns with recent findings in \cite{lcw24}, demonstrating improved convergence properties through smoother gradient updates. The training process employs decoupled weight decay regularization through the AdamW optimizer \cite{loh19}, which provides better generalization properties compared to standard stochastic gradient descent methods.

These architectural choices collectively optimize the balance between computational efficiency and decoder performance, while maintaining the theoretical foundations of our differential attention approach.

\subsection{Methodology for Comparison}
Our evaluation framework compares the proposed method, DiffMPT, with two established transformer architectures: the Error Correction Code Transformer (ECCT) \cite{chw22} and the Cross-attention Message-Passing Transformer (CrossMPT) \cite{pkk24}. We evaluate these models on Low-Density Parity Check (LDPC) codes and Polar codes, which are integral to 5G cellular network technology. The ECCT implementation is sourced directly from the authors' published codebase, while CrossMPT is implemented based on the architectural specifications in the original paper due to code unavailability.

To establish a controlled comparison environment, we standardize the architectural hyperparameters across all transformer decoders: 6 attention layers, 128-dimensional input embeddings, and 512-dimensional feedforward neural networks. This configuration ensures that all models have identical numbers of trainable parameters in their decoder layers.

The hyperparameter $\lambda_\text{multiloss}$ is set to the reciprocal of the code length ($1/n$), empirically determined to balance the magnitudes of transportation loss $\mathcal{L}_\text{transport}$ and validation loss $\mathcal{L}_\text{validation}$, ensuring comparable gradient contributions. The differential transformer's $\lambda_\text{diff}$ is implemented as a learnable parameter constrained to [0,1].

The training phase consists of 1000 epochs, each comprising 1000 mini-batches of 128 samples. Codewords $\mathbf{x}$ are uniformly sampled from the codebook, and SNR values are uniformly sampled between 2-7 dB with 1 dB intervals. We employ cosine annealing for learning rate scheduling, decreasing from $5\times10^{-4}$ to $10^{-5}$. Training stability is maintained through gradient norm clipping at 0.1 and input value clipping to [-15,15].

For performance evaluation, we generate $10^6$ random codewords per SNR value. Due to the limited number of frame errors (<100) at 6 dB and 7 dB, we focus our analysis on the 2-5 dB range with 1 dB intervals. While previous studies in the learning community predominantly report bit error rate (BER), we prioritize frame error rate (FER) reporting, as it represents a more critical metric for communication systems.

\subsection{LDPC Codes}
Given that our neural decoder design draws inspiration from message passing principles, we benchmark our performance against the Belief Propagation (BP) decoder \cite{pea14}, implementing the sum-product algorithm with 20 iterations as a strong baseline for traditional decoding approaches.

\begin{figure}[h!]
    \centering
    \subfloat[Short-to-medium codes]{
        \includegraphics[width=0.39\textwidth]{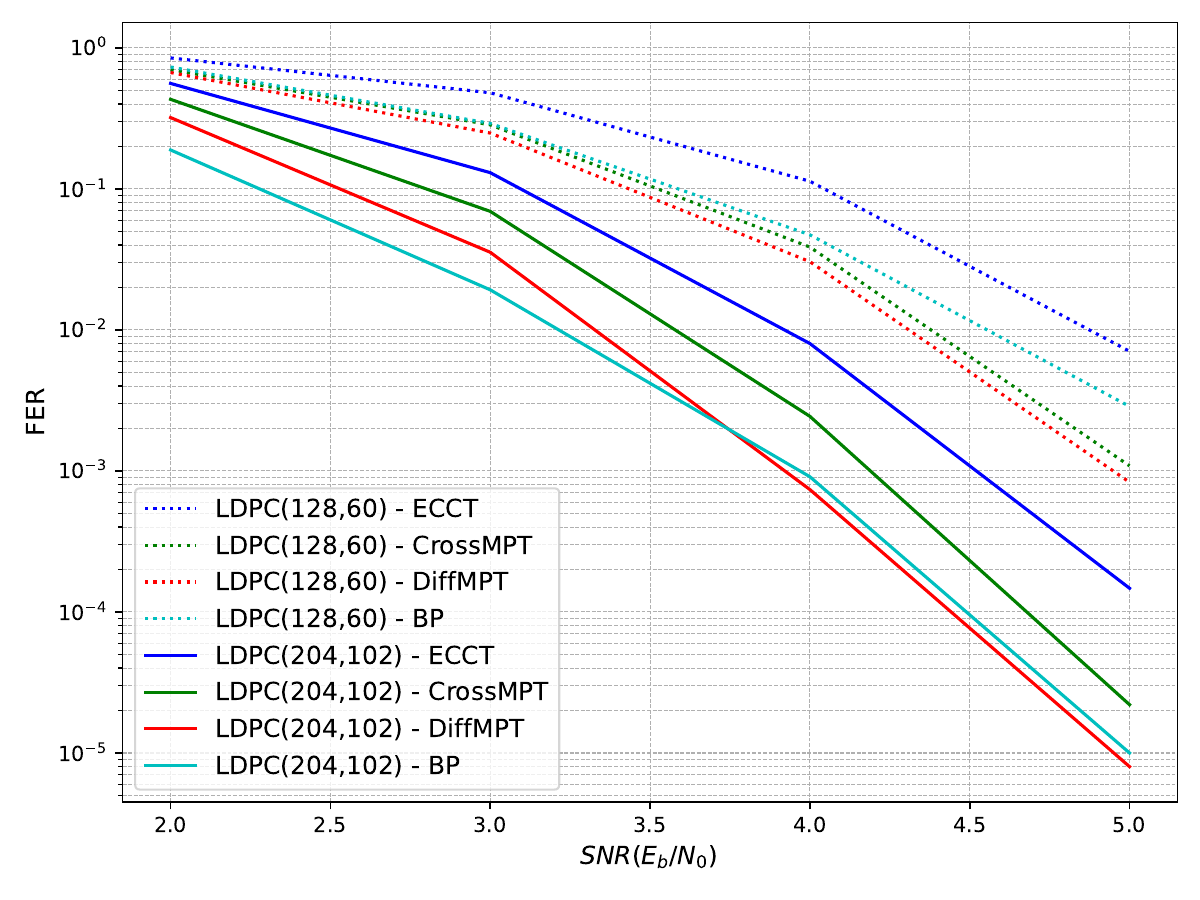}
        \label{fig:ldpc_fer1}
    }
    \hfill
    \subfloat[Long codes]{
        \includegraphics[width=0.39\textwidth]{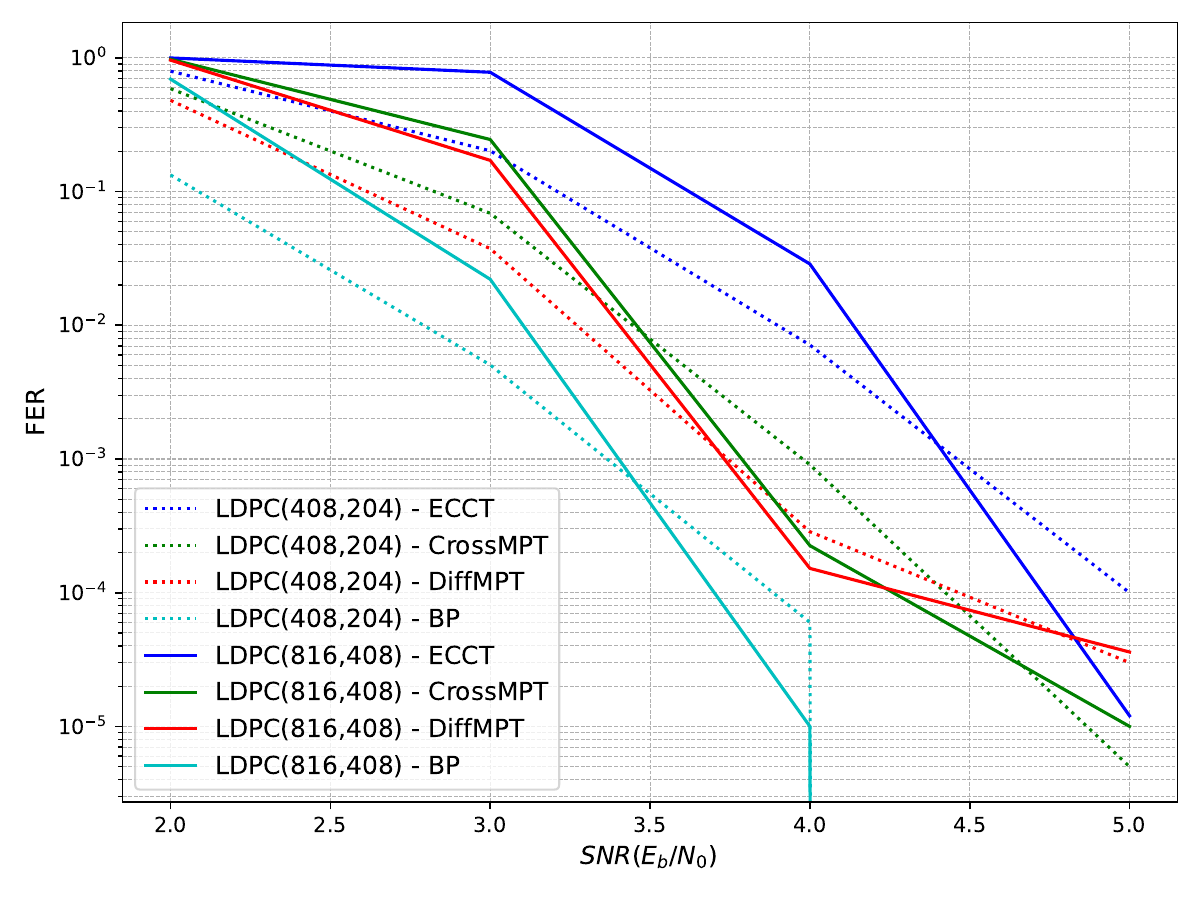}
        \label{fig:ldpc_fer2}
    }
    \caption{FER vs. SNR comparison for LDPC codes}
    \label{fig:ldpc_fer}
\end{figure}

The FER performance analysis for LDPC codes is presented in \Cref{fig:ldpc_fer}. Our model demonstrates competitive performance compared to BP decoding for short-to-medium length codes (LDPC($n = 128, k = 60$) and LPDC($n = 204, k = 102$)). Notably, for LDPC codes with length $n = 128$, our decoder achieves approximately 0.2dB gain over BP decoders at FER $10^{-2}$, highlighting the potential advantages of transformer-based architectures. These improvements stem from our efficient implementation combining differential attention mechanism with weight sharing and syndrome loss-guided training.

For longer codes (LDPC($n = 408, k = 204$) and LPDC($n = 816, k = 408$)), while our architecture maintains superior performance compared to other transformer-based approaches, all transformer decoders show a notable gap relative to BP decoders. This disparity may be attributed to the limited input embedding dimension of 128, which might be insufficient to fully capture the structural complexity of longer codes.

\subsection{Polar Code}
Given that BP decoders are suboptimal for polar code decoding, and current transformer-based approaches still show significant gaps compared to SCL decoders \cite{pkk24}, we focus our comparison on existing transformer-based architectures.

\begin{figure}[h!]
    \centering
\includegraphics[width=0.39\textwidth]{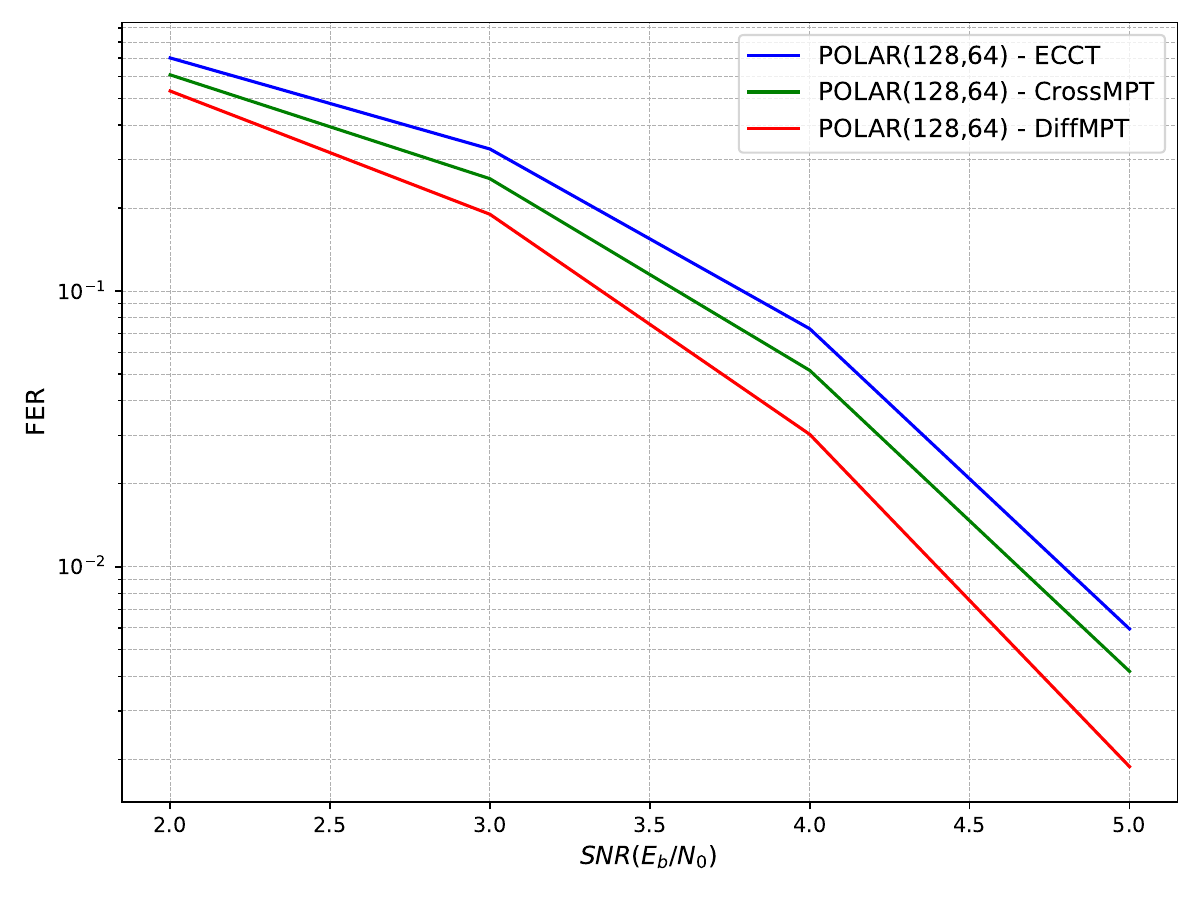}
    \caption{FER vs. SNR comparison for Polar code}
    \label{fig:polar_fer}
\end{figure}

Performance analysis shown in \Cref{fig:polar_fer} demonstrates that our architecture achieves consistent improvements over existing approaches for Polar($n=128, k=64$) codes, delivering gains of 0.2dB over CrossMPT and 0.3dB over ECCT at FER $10^{-2}$. These improvements are particularly noteworthy as they are achieved without incorporating the a priori information utilized in SCL decoders. Further performance enhancements may be possible through the integration of SCL decoder principles into our architecture.

\section{Discussion and Conclusion}
In this paper, we have introduced novel approaches to neural decoding inspired by message passing principles, notably incorporating a syndrome-based loss function and differential-attention architecture. Our experimental results demonstrate consistent FER improvements over existing transformer-based decoders. However, significant opportunities remain for further advancement by leveraging traditional coding theory.

Several promising research directions emerge from this work. First, the incorporation of codebook structural information beyond Tanner graph representations could improve decoder performance. Second, the challenge of achieving superior performance compared to BP decoders while maintaining compact latent representations remains an open problem. Addressing these challenges could bridge the remaining gap between neural and traditional decoders, particularly for longer codes.
\clearpage
\bibliographystyle{IEEEtran}
\bibliography{mybiblio}

\end{document}